\documentclass[runningheads,a4paper]{llncs}

\usepackage{url}
\usepackage{graphicx}	
\usepackage{color, colortbl}
\usepackage{amssymb}

\usepackage{multirow}

\usepackage{listings}
\usepackage{soul} 
\usepackage{color}
\definecolor{KWColor}{rgb}{0.37,0.08,0.25}
\definecolor{CommentColor}{rgb}{0.133,0.545,0.133}
\definecolor{StringColor}{rgb}{0,0.126,0.941}
\usepackage{balance}

\newcommand{\tool}[1]{{\textbf{AndroZoo}}}

\title{AndroZoo++: Collecting Millions of Android Apps and Their Metadata for the Research Community}
\titlerunning{AndroZoo++}

\author{Li Li, Jun Gao, M\'ed\'eric Hurier, Pingfan Kong, Tegawend\'e F. Bissyand\'e, 
Alexandre Bartel, Jacques Klein, Yves Le Traon}
\authorrunning{Li Li et al.}

\institute{
Interdisciplinary Centre for Security, Reliability and Trust (SnT),\\ 
University of Luxembourg, Luxembourg\\
\email{firstName.lastName@uni.lu}\\
}

\begin{document}

\maketitle

\begin{abstract}

We present a growing collection of Android apps collected from several sources, including the official Google Play app market and a growing collection of various metadata of those collected apps aiming at facilitating the Android-relevant research works.
Our dataset by far has collected over five million apps and over 20 types of metadata such as VirusTotal reports.
Our objective of collecting this dataset is to contribute to ongoing research efforts, as well as to enable new potential research topics on Android Apps. 
By releasing our app and metadata set to the research community, we also aim at encouraging our fellow researchers to engage in reproducible experiments.
    
This article will be continuously updated based on the growing apps and metadata collected in the AndroZoo project.
If you have specific metadata that you want to collect from AndroZoo and which are not yet provided by far, please let us know.
We will thereby prioritise it in our collecting process so as to provide it to our fellow researchers in a short manner.
    
\end{abstract}

\section{Introduction}

Mobile app development has witnessed an unprecedented growth in recent years due to the increase in affordability and adoption of smart powerful handheld devices. In particular, the Android ecosystem, with its open Operating System and the available Software Development Kit, have empowered developers to produce millions of apps for diverse user tasks, ranging from emails and games to payment and health activities.
Unlike the few established traditional desktop applications which have been thoroughly studied by the research community, Android apps are legion, each having a large share of user base. Analysing these apps at a large scale is however challenging since market maintainers implement several restrictions in collecting apps. In this context, researchers proceed in the best effort way to reuse small datasets (which are generally obsolete) or collect a limited number of samples (which may not be representative), leading to studies which may be biased and experiments which are often not reproducible.
To address the problem of Android dataset collection, we have invested in a long-term effort to crawl apps for the research community. After several years of crawling, we have already stored over five million of apps. With AndroZoo, we aim to provide the software engineering research community with an unrestricted, scalable and up-to-date access to Android apps. To that end, we have developed specialised crawlers for several market places to automatically browse their content, find Android applications that could be retrieved for free, and download them into our repository.
To the best of our knowledge, the total number of apps that we have collected constitutes the largest dataset of Android apps ever used in published Android research studies.
Often, it is impossible to know beforehand how many apps are available in a given market. Therefore, some of the markets for which we wrote dedicated crawlers proved to be much smaller than initially expected.
The crawlers we wrote follow two main objectives: 
a) Collect as many apps as possible, and 
b) Ensure the lowest possible impact on the market infrastructure. 
These two objectives increased the cost of writing such crawlers since for every market a manual analysis of the website has been performed in order to detect and filter out pages with different URL but with similar contents, for example, lists that can be sorted according to different criteria. Similarly, a unique identifier for every APK on one market had to be found, so that deduplication can happen before downloading apps.
While reducing the load we incur to markets’ web servers may not seem strictly necessary to the objective of collecting apps, it vastly reduces the likelihood of being banned by market owners and hence, helps building and maintaining in the long term a large and up-to-date dataset.

This article is an extended version of a data paper entitled ``AndroZoo: Collecting Millions of Android Apps for the Research Community'' published at the International Conference on Mining Software Repositories (MSR) 2016.
In this extension, in addition to collect Android apps for the research community, we further provide to the research community various metadata of Android apps.
The main idea behinds this extension is to facilitate the access of our AndroZoo app set.
Indeed, based on the feedback collected from AndroZoo's users, we observe that the main bottleneck of analyzing the whole AndroZoo app set is the bandwidth.
A lot of time and resources are needed in order to access all the available AndroZoo apps.
However, since some apps in AndroZoo may not be necessary for certain analysis (e.g., sending apps that have no reflection access to reflection analysis tools), 
it could be much efficient to only download and analyse (the small set of) relevant apps.
To this end, we provide to the research community various metadata of Android apps that are expected to be used by users of \tool{} to pre-select relevant apps.
Furthermore, since it is usually unavoidable to compute the metadata of Android apps for supporting further analysis, our effort, which is done once, could potentially avoid the same work being repeated by other researchers, resulting in less computing resources spent in our society.

\textbf{Outline.}
The remainder of this article is organised as follows:
Section~\ref{sec:app_collection} presents the architecture developed to collect Android apps while
Section~\ref{sec:metadata_collection} presents the collection of various metadata associated to \tool{} apps.
Section~\ref{sec:implications} then enumerates some implications that leverage \tool{} to perform various analyses.
Finally, Section~\ref{sec:leveraging} and~\ref{sec:conclusion} discusses some limitations and provides concluding remarks respectively.

The AndroZoo apps and metadata are available at the following website:
\begin{center}
\url{https://androzoo.uni.lu}
\end{center}

\section{App Collection}
\label{sec:app_collection}

For most app sources, we developed a dedicated web crawler, where every candidate app which is available for free runs through a processing pipeline that:
(1) Ensures this app has not already been downloaded; 
(2) Downloads the file;
(3) Computes its SHA256 checksum;
(4) Archives the file.
To check that an app has not been already downloaded, we first identify a unique identifier for APKs in the market associated to the crawler, and store in a database an entry market name–app identifier. 
As a consequence, and because it is impossible to determine that two files from two markets are the same unless both are downloaded and compared, the deduplication is local to one market, meaning that one file from one market is downloaded exactly once, regardless of whether or not it has already been downloaded from another market.

By far, we are collecting apps from around ten app markets or repositories.
We now describe some representative ones

\begin{itemize}

\item Google Play. The official market of Android is a website that allows users to browse its content through a web browser. Apps cannot however be downloaded through a web browser. Instead, Google provides an Android app2 that uses a proprietary protocol to communicate with Google Play servers. No app, however, can be downloaded from Google Play without a valid Google account (not even free Apps). Both issues thus outlined were overcome using open-source implementations of the proprietary protocol and by creating free Google accounts. The remaining constraint was time, as Google also enforces a strict account-level rate-limit: a given account is not allowed to download more than a certain number of apps in a given time frame.
Google Play has several features that make automatic crawling harder than other markets. As a result, a more elaborated crawler is required for this market. Google further enforces limits on the number of apps that can be downloaded per Google account in a given period from one IP address.
To overcome those limits, we wrote a software dedicated to finding and downloading apps from Google Play. This software is built with two components: a central dispatcher, and a download agent.
We have used agents on up to seven machines located in Luxembourg, France and Canada. On three of these machines, we ran two instances of the agent, one using exclusively IPv4 connectivity and the other using IPv6. Because IPv4 and IPv6 addresses are not linked in any way, this allows to hiding the fact that those two agents run on the same machine, hence enabling us to increase the number of apps downloaded from one computer without increasing the risk of being blacklisted.

\item Anzhi. Anzhi, the largest alternative market of our dataset, is operated from China and targets the Chinese Android user base. It stores and distributes apps that are written in the Chinese languages, and provides a less strict screening policy than that of Google Play.

\item AppChina. AppChina, another Chinese market, has used to enforce drastic scraping protections such as a 1Mb/s bandwidth limitation and a several-hour ban if using simultaneously more than one connection to the service.

\item F-Droid.
F-Droid is a repository of Free and open-source Android apps that users can download and install on their devices. 
Many of the apps found on F-Droid are modified versions of apps that are released to other markets by their developers. The modifications brought by F-Droid are usually linked for example with advertisement.

\end{itemize}

\section{Metadata Collection}
\label{sec:metadata_collection}

Our objective in this work is to facilitate the access of AndroZoo by providing to the research community not only a large set of Android apps but also sufficient metadata associated with those apps.
Users of AndroZoo can therefore to pre-select a set of apps that are all relevant to their research.
As a result, only those selected apps need to be downloaded, resulting in fewer constraints to AndroZoo's users, who otherwise have to download all the available apps of AndroZoo, confronting massive resource costs (e.g., bandwidth and disk).

Table~\ref{tab:overview} presents an overview of the metadata we currently provide.
So far, we have collected for the research community four types of metadata containing in total 24 items.
We now provide brief descriptions to them respectively.

\begin{table*}[!t]
\centering
\caption{Metadata Overview.}
\label{tab:overview}
\resizebox{\linewidth}{!}{
\begin{tabular} { l l l }
\hline
Type    & Name              & Example \\

\hline

\multirow{6}{*}{APK}    
    & SHA256 & 00010F599EB0EDC52F5781E1892B7B50B922055FCDF766ECC45B50C8E25E84B6          \\
    & SHA1              & 9AC02C1AACF6A7E41E013E19677DDEBD1DC7900D           \\
    & MD5               & C20F5F99F20745109591DD422C4DCF56          \\
    & Apk Size         & 8156810          \\
    & Market            & play.google.com \\
    & Certificate      & C=RU, ST=MO, L=Moscow, O=gera, OU=gera, CN=gera           \\

\hline

\multirow{5}{*}{Manifest} 
 & Application ID         & com.gera.ArtScienceMuseumJigsawPuzzles \\
 & Version Code         &  1     \\
 & API Level            &  14     \\
 & Permission List      &  cf. Listing~\ref{code:perm_list}     \\
 & Feature List         &  cf. Listing~\ref{code:feature_list} \\

\hline

\multirow{7}{*}{DEX}     
     &   Dex Size         &  8247212     \\
     &   Dex Date    &  14/02/2016  00:36:34     \\
     &   Native Code    & true      \\
     &   Crypto Code    & true      \\
     &   Dynamic Code    & true       \\
     &   Reflection    & true       \\
     &   Class List  &  cf. Listing~\ref{code:class_list}     \\

\hline

\multirow{8}{*}{Releasing$^\alpha$}
    &   Description & Art Science Museum Puzzle Jigsaw game dedicated to ...   \\
    &   Category    & Puzzle \\
    &   Author      & gera \\
    &   Rating Score & 4.0  \\
    &   Installs    &   100 $-$ 500 \\
    &   Requires Android & 4.0 and up \\
    &   Updated Date & May 13, 2016\\
    &   Contact   & Email geraraaaa@gmail.com \\
    
\hline

\multirow{2}{*}{Security} 
 & VirusTotal Report    & cf. Listing~\ref{code:virustotal_report}   \\
 & AndroBugs Report     & cf. Listing~\ref{code:androbugs_report}  \\

\hline

\multirow{4}{*}{Others} 
 & Piggybacking App Pairs & \url{https://github.com/lilicoding/Piggybacking}   \\
 & Common Libraries & \url{https://github.com/lilicoding/CommonLibraries/blob/master/libraries/cl_91.txt}  \\
 & Ad Libraries     & \url{https://github.com/lilicoding/CommonLibraries/blob/master/libraries/ad_240.txt}   \\
 & App Lineages & cf. Listing~\ref{code:app_lineage}  \\

\hline

\multicolumn{3}{l}{$^\alpha$ Example\ presented\ in\ this\ type\ is\ based\ on\ the\ latest\ information\ of\ Google\ Play.} \\

\end{tabular}}
\end{table*}

\subsection{APK}
Android apps are released and distributed through APK files.
For each Android APK file, we provide six artefacts for our fellow researchers.

\begin{itemize}

\item \textbf{SHA256, SHA1, MD5:}
We provide three hash values (SHA256, SHA1, and MD5) to uniquely identify an Android app.

\item \textbf{APK Size:}
We provide the APK size information (in byte) of a given app aiming at providing a quick means for researchers to calculate how much space they may need in order to download the app from \tool{} to their local disk.

\item \textbf{Market:}
For each APK, when crawling from a market, we also record from which market it is downloaded.
Since we do not have a good means to check if a given APK has already been downloaded from other markets, the same APK could be downloaded from several markets.
This market artefact represents the market where the app is downloaded from.
If the app appears in multiple markets, the market artefact correspondingly shows all the available markets with each separated by a vertical virgule (e.g., \emph{play.google.com|anzhi} indicates that the app is downloaded from both Google Play and the Anzhi market).

\item \textbf{Certificate:}
we provide the certificate signature of a given app, which is usually used to represent the signature of its developers.
Literature works recurrently leverage this metadata to pinpoint repackaged (or piggybacked) Android apps, which usually involve a change of app signatures (i.e., developers)~\cite{li2017understanding}.

\end{itemize}

\subsection{Manifest}
Every Android app is assembled with a global configuration file called \emph{AndroidManifest.xml} (hereinafter referred as manifest), which plays an important role for configuring app's permissions, components, etc.
In this work, we provide five artefacts that are directly extracted from the manifest file of a given app.
Those five artefacts are as follows:

\begin{itemize}
\item \textbf{Application ID:} 
Application ID is used to uniquely identify an app on a device and in Google Play.
For example, two apps with the same application ID cannot be installed concurrently on the same device.
It is also disallowed to update an app in Google Play with a different application ID.

\item \textbf{Version Code:} 
Version code is provided for app developers to name the different app versions.
Theoretically, version code increases as the app updated.
Literature work leverage this information to build app lineages (i.e., sorted app versions of the same app) for supporting the investigation of app evolution.

\item \textbf{API Level:}
API level specifies the SDK version that the app is implemented.
Since different API levels may provide different features, users of \tool{} could thus leverage this information to conduct evolution-based investigations, especially in couple with the evolution investigation of the Android operating system.
As an example, Li et al.~\cite{li2016accessing} have leveraged this information to investigate the evolution of inaccessible Android APIs (i.e., internal and hidden APIs) of the framework code.

\item \textbf{Permission List:}
Permissions, declared by app developers, are used by Android system to grant the access of protected parts and controls the interaction with other apps. 
Listing~\ref{code:perm_list} illustrates an example of permission list.
Many state-of-the-art works have leveraged permissions to perform specific analyses~\cite{felt2011android,backes2016demystifying}.
In this work, we provide directly the used permissions to the research community for boosting the research of permission-based investigations.
Uses of \tool{} now can collect the declared permissions of a given app without actually downloading the app.

\item \textbf{Feature List:}
Similar to the permission list, we provide the feature list containing features defined in the manifest (cf. Listing~\ref{code:feature_list}) to the research community for helping researchers pre-select such apps that have all used certain features.

\end{itemize}

\begin{lstlisting}[
caption={Permission List Example. This List is Extracted from App \emph{com.gera.ArtScienceMuseumJigsawPuzzles}.},label=code:perm_list,float,firstnumber=1]
android.permission.ACCESS_NETWORK_STATE
android.permission.INTERNET
android.permission.ACCESS_COARSE_LOCATION
android.permission.WRITE_EXTERNAL_STORAGE
android.permission.ACCESS_WIFI_STATE
android.permission.SYSTEM_ALERT_WINDOW
android.permission.SET_WALLPAPER
android.permission.READ_EXTERNAL_STORAGE
\end{lstlisting}

\begin{lstlisting}[
caption={Feature List Example. This List is Extracted from App \emph{com.gera.ArtScienceMuseumJigsawPuzzles}.},label=code:feature_list,float,firstnumber=1]
android.hardware.faketouch
android.hardware.location
android.hardware.screen.portrait
android.hardware.wifi
\end{lstlisting}

\subsection{DEX}
DEX is the file format of the so-called Dalvik Executable format, which stores the actual code of Android apps.
Every Android app should have a DEX file called \emph{classes.dex} in the top direct of a given APK file, which is also the target where all the metadata in this group collected from.

\begin{itemize}
\item \textbf{Dex Size:} 
Similar to APK size, Dex size provides another means for users of \tool{} to pre-select their favoured apps, e.g., apps with their Dex size smaller than 1 megabyte.

\item \textbf{Dex Date:} 
Dex (assembly) date is extracted based on the last modified time of the Dex file.
It can be used to represent the assembling time of a given app and thereby to support all the time-relevant investigations.

\item \textbf{Native Code:}
Native code, if true, shows that the app has accessed native code, which is usually written in C or C++.

\item \textbf{Crypto Code:}
Crypto code, if true, reveals that the app has somehow leveraged crypto code.
Researchers such as the authors of~\cite{egele2013empirical} could benefit from this artefact to only collect (e.g., download from \tool{}) crypto-relevant Android apps, avoiding the download of irrelevant apps and also the analysis of those potential irrelevant apps.

\item \textbf{Dynamic Code:}
Dynamic code, if true, indicates that the app has loaded additional code at runtime.

\item \textbf{Reflection:}
Reflection, if true, demonstrates that the app has accessed reflective code.
Usually, if the dynamic code item is true, reflection item should be also true, as the additionally loaded code should normally be accessed through reflection.

\item \textbf{Class List:}
Class list enumerates all the class names of an app (cf. Listing~\ref{code:class_list}).
We expect this information to be used by researchers for selecting apps with accessing specific classes.
For example, researchers could leverage this information to select a set of apps with which all of them have integrated with ad library \emph{com.wandoujia.ads}.

\end{itemize}

\begin{lstlisting}[
caption={Simplified Example of Class List.},label=code:class_list,float,firstnumber=1]
com.amazon.device.ads.Ad
com.amazon.device.ads.AdActivity
com.applovin.impl.sdk.cg
com.applovin.impl.sdk.ch
com.appodeal.ads.interstitial.v
com.appodeal.ads.interstitial.w
com.chartboost.sdk.CBLocation
com.chartboost.sdk.Chartboost
com.facebook.ssp.internal.g$7
com.facebook.ssp.internal.g$a
com.gera.ArtScienceMuseumJigsawPuzzles.BuildConfig
com.gera.ArtScienceMuseumJigsawPuzzles.R
com.google.ads.AdRequest$Gender
com.google.ads.AdSize
com.google.android.gms.internal.zzcy
com.google.android.gms.internal.zzcz
com.inmobi.rendering.InMobiAdActivity$4
com.inmobi.rendering.InMobiAdActivity$5
com.mopub.common.AdFormat
com.mopub.common.AdReport
com.revmob.android.d
com.revmob.android.e
com.startapp.android.publish.Ad
com.startapp.android.publish.Ad$1
com.yandex.metrica.a
com.yandex.metrica.b
ru.mail.android.mytarget.core.enums.b
ru.mail.android.mytarget.core.facades.a
scala.collection.SeqView
scala.collection.SeqViewLike
\end{lstlisting}

\subsection{Releasing}
In addition to APK-related metadata, we also attempt to collect other metadata that are published along with the release of the app, e.g., such information that is provided by Android app markets such as Google Play.
So far, we collect eight artefacts for an APK and the collection is only performed for Google Play apps.
It is also worth to mention that we only store the latest releasing metadata.
In other words, based on our current policy, even if the releasing metadata exists for a given app, our crawling program will overwrite them anyway based on the newly crawled information.

\begin{itemize}

\item \textbf{Description:}
Description artefact is a paragraph of text that is provided by app developers to depict the basic functionality of the app, so as to motivate users to download and install.
Researchers could leverage this information to check against the app code behaviour, as what has been presented by Gorla et al.~\cite{gorla2014checking}.

\item \textbf{Category:}
Category artefact illustrates the group that the app belongs to.
Normally, apps from the same group would potentially share more or less the same functionality.
Researchers could leverage this information to demonstrate the diversity of their app set.
In general, more category considered, more diverse the dataset is.

\item \textbf{Author:}
Author artefact indicates the developer of the app.

\item \textbf{Rating Score:}
Rating Score artefact is a float number that is computed based on all the app reviewers' ratings, which is also used by app markets to rank their hosted apps.

\item \textbf{Installs:}
Installs artefact illustrates the range of possible installs the app has received.
Roughly speaking, the more installs an app receives, the more popular the app is.

\item \textbf{Requires Android:}
Requires Android artefact provides the range of Android OS versions that the app requires running.
For example, \emph{4.0 and up} indicates that the app can only be installed and launched at devices running Android OS with version \emph{Ice Cream Sandwich} (i.e., 4.0 or API level 14) or above versions that are released later.

\item \textbf{Updated Date:}
Updated Date artefact illustrates when the latest app version is uploaded to the market.

\item \textbf{Contact:}
Contact artefact provides the email address of the authors of the app.
Researchers could resort to this information to contact the app developers for recommending bug fixes or simply requesting for answering some surveys.

\end{itemize}

\subsection{Security}
As revealed in our recent systematic literature review (SLR), security remains the most targeted topic in the field of static analysis of Android apps~\cite{li2017static}.
Towards boosting security analysis of Android apps, in this work, we also collect several security reports from well-known tools that we share to the research community.

\begin{itemize}
\item \textbf{VirusTotal Report:} 
VirusTotal\footnote{https://virustotal.com} is a free service that leverages over 50 anti-virus products to analyze suspicious files including Android apps.
We have sent all the \tool{} apps to VirusTotal and consequently collected the security reports for all of those apps.
Listing~\ref{code:virustotal_report} illustrates a simplified example of VirusTotal report, where four anti-virus products (i.e., Bkav, ESET-NOD32, Fortinet, and AVG) have flagged app \emph{00010F} as malicious while the remaining anti-virus products such as McAfee, Qihoo-360 do not flag it as such.

\item \textbf{AndroBugs Report:} 
AndroBugs\footnote{https://www.androbugs.com} is an Android vulnerability scanner that assists developers or hackers find potential security vulnerabilities in Android apps.
Listing~\ref{code:androbugs_report} presents a simplified example of AndroBugs report, which not only discloses the critical vulnerability problems but also reveals the exact point where the vulnerability appears.

\end{itemize}

In a near future, we plan to also include the security reports of other well-known tools such as FlowDroid and IC3.

\begin{lstlisting}[
caption={Simplified Example of VirusTotal Report.},label=code:virustotal_report,float,firstnumber=1]
{"_id":"00010F599EB0EDC52F5781E1892B7B50B922055FCDF766ECC45B50C8E25E84B6",
"scans":{
"Bkav":{"detected":true,"result":"Android.Adware.RevMob.88C5"},
"ESET-NOD32":{"detected":true,"result":"a variant of Android/Inmobi.C potentially unsafe"},
"Fortinet":{"detected":true,"result":"Android/Generic.S.1BE3EB!tr"},
"AVG":{"detected":true,"result":"Android/G2P.R.0C64D4E32B5E"}
"McAfee":{"detected":false,"result":null},
... ...
"Qihoo-360":{"detected":false,"result":null}
}}
\end{lstlisting}

\begin{lstlisting}[
caption={Simplified Example of AndroBugs Report.},label=code:androbugs_report,float,firstnumber=1]
[Critical]  App Sandbox Permission Checking:
 [getSharedPreferences] Lcom/google/android/gms/flags/impl/zzb$1;->zzvw()Landroid/content/SharedPreferences;
 [openFileOutput] Lcom/mopub/mobileads/AdAlertReporter;->addImageAttachment(Ljava/lang/String; Landroid/graphics/Bitmap;)V
[Critical] <Implicit_Intent> Implicit Service Checking:
 com.yandex.metrica.MetricaService
[Critical] <SSL_Security> SSL Connection Checking:
http://a.applovin.com/
 Lcom/applovin/impl/sdk/bw;-><clinit>()V
[Critical] <WebView><Remote Code Execution><#CVE-2013-4710#> WebView RCE Vulnerability Checking:
 Lcom/startapp/android/publish/slider/b;->b()V --->
 Landroid/webkit/WebView;->addJavascriptInterface(Ljava/lang/Object; Ljava/lang/String;)V
\end{lstlisting}

\subsection{Others}

In addition to the aforementioned metadata, in this work, we also collect several advanced ones that are collected following specific rules.

\begin{itemize}
\item \textbf{Piggybacking App Pairs:} 
Piggybacking constitutes a specific subset of repackaging, is defined in the literature as an activity where a given Android app is repackaged after manipulating the app content, e.g., to insert a malicious payload, an advertisement library, etc.
In this work, we provide to the research community a set of piggybacking pairs that are collected based on the following rules:
1) The original app is benign while the piggybacked app is malicious.
2) The two apps share the same application ID.
3) The two apps are signed by different signatures (one must not be the real one).
4) The two apps are developed based on the same SDK version.
5) The two apps have over 80\% of similarity in terms of app code.

\item \textbf{Common/Ad Libraries:}
Despite some efforts on investigating Android libraries, the momentum of Android research has not yet produced a complete set of common libraries to further support in-depth analysis of Android apps.
We, therefore, leverage a dataset of about 1.5 million apps from Google Play to harvest potential common/ad libraries. 
With several steps of refinements, we finally collect a set of 1,113 libraries supporting common functionality and a set of 240 libraries for advertisement.

\item \textbf{App Lineages:}
App lineages, which represent the evolution history of a given app (cf. Listing~\ref{code:app_lineage} illustrates an app lineage example), is valuable for various research directions.
For example, they can be leveraged in practice to infer API usage updates that app developers adopt with respect to API changes in the framework.
They can also be used to infer bug fix patterns and vulnerability repair schemes.
Our main objective in this work is to present to the community a large set of app lineages that can later be leveraged by researchers and practitioners to support or facilitate their analyses.

\end{itemize}

\begin{lstlisting}[
caption={An App Lineage Example (cf. App \emph{es.ligabbva.gi.main}).},label=code:app_lineage,float,firstnumber=1]
%SHA256(VERSION_CODE)
134D13BC666BD0AC04F052BABAF024257E92C4AC4484FF701EF485CFB64B3B19(118)
7679DFC9BF79F80517E038FE1177D810A888822E2A569143B4234ADA7BE60928(120)
42E9FF2BADF5A1A1B9ABDCE38633E18F94D8D448E7C7EE63661C6086F190EC49(121)
B3C32223F3095AB23FD0E5B1776FFD8A49DAD3E19756EC96EC356371B7248863(122)
A26A6424E243C5AD1AD1E4D564ED109A5D6D5D569560166852FD7EE65CCB4C80(123)
C1DE441C3DC1DE7901062ECC31C92770F24FBF5DFDF1A5A8659B3DBCEF9EDF92(130)
3104705FFC077F677C06149FECA8C9090AA1CA1E2289808B1EAFB1FF27DAD31F(133)
ACF850967DA85BED5EA646E5A16AF5E6B2DD914F79D908A33E57C1CE91D034F3(134)
1164D359138EEB00621C2BE75B7F86FAB0A1AE53840A14AF763487458A0F4BFD(135)
63709D29B632BBCF213F2DE86FAA504076A7ECFC44694BE4D626F38F0AB01A94(136)
B282170AFAE0A78605FD1EF71C4C46477B1B727F2F3F8B7120F11EB5DEB8172F(138)
B2A00A677B7A480416349EEA394D4F9003E7F1CCE05792687B5357F0B5D147AA(141)
9F45A1EA7EFBE29E00C7D6CC1F9B8993C62720DA46FD2E8077E435733D235E32(142)
EFB36CDF51C0721A5DF9753D43106ECD64D22E8CAA0ED5A20135A0F104869F3B(145)
296F4FA58F16669A2A1B853F1FE55D7E7245B3F41B1DE394D96E41F7A9BF505D(151)
1DCEB8962D345F674A4DC83D0165D5A1816F310FCD9AD63E1BE6A591B2D21A89(162)
C302071F4D2362C9C784E8704428A30E8EAE2B4BA3F2954F8C9D4E33ED7583DE(163)
6B377D3625DF8AAC6AD066550BB110F4D75F68BC98D15B697CD8A6E70EA464E5(165)
\end{lstlisting}



\section{Implications}
\label{sec:implications}

\begin{figure}[!h]
    \centering
    \includegraphics[width=0.8\linewidth]{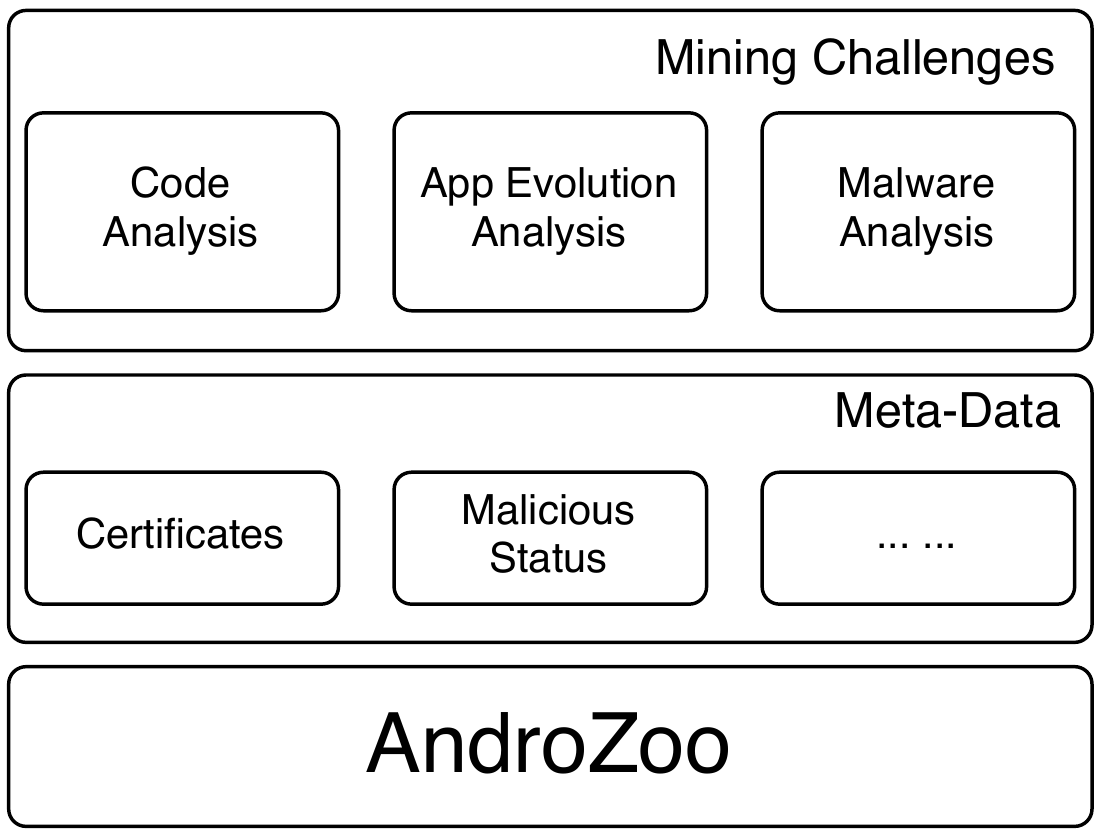}
    \caption{Overview of Our Research Dimensions.}
    \label{fig:overview}
\end{figure}

Fig.~\ref{fig:overview} presents an overview of possible implications that we have explored so far based on the AndroZoo apps and their meta-data that we pre-compute for facilitating high-level analyses.
In a nutshell, the implications on mining Android apps are mainly located in three dimensions:
(1) Code analysis; 
(2) App evolution analysis; and
(3) Malware analysis;
We now detail those research dimensions respectively.

\subsection{Code Analysis}

\textbf{ICC-Aware Analysis.}
Since Android apps can leak sensitive information carelessly or maliciously and malicious apps manipulate significantly more ICCs than benign apps, we propose a static analysis approach named IccTA to detect privacy leaks crossing Android components, and crossing apps with the help of ApkCombiner~\cite{li2015apkcombiner}.
Unlike state-of-the-art approaches, which mainly detect privacy leaks within single component, 
IccTA propagates context information among components to support inter-component communication (ICC) analysis~\cite{li2015iccta,li2014know}.

IccTA applies a code instrumentation based approach, i.e., change the code before analyzing, in order to make the ICC analysis reusable to existing intra-component analyzers.
Listing~\ref{code:example_icc} illustrates a code snippet showing an ICC-based privacy leak.
The device id, considered as sensitive information, is obtained and stored into an Intent object (lines 8-10) in Activity1.
Then, an ICC method \emph{startActivity} is called, which switches the current execution from Activity1 to Activity2.
Finally, in Activity2, the device id is retrieved and is eventually sent out of the device through \emph{sendTextMessage} (lines 15-17).

\begin{lstlisting}[caption={Example of an ICC Leak.},label=code:example_icc,float=h]
//TelephonyManager telMnger; (default)
//SmsManager sms; (default)
class Activity1 extends Activity {
 void onCreate(Bundle state) {
  Button to2 = (Button) findViewById(to2a);
  to2.setOnClickListener(new OnClickListener(){
   void onClick(View v) {
    String id = telMnger.getDeviceId();
    Intent i = new Intent(Activity1.this,Activity2.class);
    i.putExtra("sensitive", id);
    Activity1.this.startActivity(i);
}});}}
class Activity2 extends Activity {
 void onStart() {
  Intent i = getIntent();
  String s = i.getStringExtra("sensitive");
  sms.sendTextMessage(number,null,s,null,null);
}}
\end{lstlisting}

This privacy leak cannot be detected by intra-component analyzers such as FlowDroid~\cite{arzt2014flowdroid}, because the switching between Activity1 and Activity2 is unfortunately decided only by the system and it is non-trivial to obtain it directly at the code level~\cite{octeau2015composite,octeau2016combining}.
Therefore, in this work, we present IccTA, a code instrumentation based approach, which modifies the code to be analyzed in a way that inter-component feature is mitigated.
As an example, Listing~\ref{code:resolve_icc} demonstrates the modifications made by IccTA for the ICC leak example shown in Listing~\ref{code:example_icc}.
The ICC method \emph{startActivity} is replaced by a helper method that simulates the ICC through Java code, resulting in a simplified code snippet where ICC is no longer appearing.
As a consequence, existing intra-component analyzers such as FlowDroid can now detect the privacy leak shown in Listing~\ref{code:example_icc} without any modification.

\begin{lstlisting}[caption={Code Instrumentation for \emph{startActivity}.},label=code:resolve_icc,float=h]
  // modifications of Activity1
- Activity1.this.startActivity(i);
+ IpcSC.redirect0(i);
  // creation of a helper class
+ class IpcSC {
+  static void redirect0(Intent i) {
+   Activity2 a2 = new Activity2(i);
+   a2.dummyMain();
+  }
+ }
  // modifications in Activity2
+ public void dummyMain() {
+  // lifecycle and callbacks
+  // are called here 
+ }
+ public Activity2(Intent i) {
+   this.intent_for_ipc = i; }
+ public Intent getIntent() {
+  return this.intent_for_ipc; }
\end{lstlisting}

\textbf{Reflection-Aware Analysis.}
Like ICC we introduced previously, reflection is another challenge that usually causes static analyzers to yield false negatives.
Android developers heavily use reflection in their apps for legitimate reasons such as providing genericity, maintaining backward compatibility, accessing inaccessible APIs~\cite{li2016accessing}, but also significantly for hiding malicious actions.
Unfortunately, based on our SLR~\cite{li2017static}, most state-of-the-art works do not take into account the presence of reflective calls.
Hence, we present DroidRA~\cite{li2016droidra,li2016reflection}, a code instrumentation based approach, to tackle this issue in a non-invasive way.
We first model the reflection analysis problem to a constant string propagation problem and then leverage the COAL solver~\cite{octeau2015composite} to infer the values of reflection-related targets.
Finally, we instrument the app code to replace reflective calls by traditional Java calls where the separated parts due to reflection are now connected.
Experimental results demonstrate that DroidRA is capable of supporting state-of-the-art static approaches to provide more sound and complete analyses.

\textbf{Harvesting Parameter Values of Android APIs.}
Parameter values are important elements for understanding the usage of Application Programming Interfaces (APIs) of certain Android APIs.
For example, Android malware may assign parameter values to certain APIs that are, statistically speaking, different from that assigned by benign apps.
Furthermore, not only for identifying security issues, understanding the usage of parameter values could also benefit app developers by recommending to them possible inputs for certain APIs.
Therefore, there is a strong need to investigate the parameter values of Android apps.
In this work, we achieve this purpose by performing static code analysis, where we introduce a prototype tool named ParamHarver to automatically extract API parameter values from Android apps~\cite{li2016parameter}.
Our experimental investigations and findings further illustrate that a thorough study of parameter values could be leveraged in various scenarios.

\textbf{Common Library Analysis.}
In a preliminary study, Wang et al.~\cite{wang2015wukong} have found that over 60\% of Android apps' code is from common libraries, which may not be relevant to certain analyses such as repackaging analysis.
Indeed, there are a number of research work that excludes libraries before performing their analyses~\cite{avdiienko2015mining,gibler2013adrob,grace2012unsafe}.
Despite some efforts on investigating common libraries, the momentum of Android research has not yet produced a complete set of libraries that can be taken as a whitelist to support further analysis.
Therefore, in this work, we leverage AndroZoo apps to perform a heuristic-based approach with several steps of refinements to harvest potential common libraries, for which we eventually collect 1,113 general libraries and 240 advertisement libraries.
Furthermore, based on the collected libraries, we have performed several empirical studies that confirm our motivation: certain analyses such as repackaging analysis and malware detection should not take into account the library code, which constitutes noise in app features, in order to produce accurate results.

\subsection{App Evolution Analysis}

\textbf{App Variant Analysis.}
App variants (or family, e.g., the different products of the same company) usually embrace valuable information for evaluating extractive Software Product Line (SPL) adoption techniques, e.g., the reuse practices among app variants~\cite{martinez2017bottom}.

\begin{figure}[!h]
    \centering
    \includegraphics[width=0.8\linewidth]{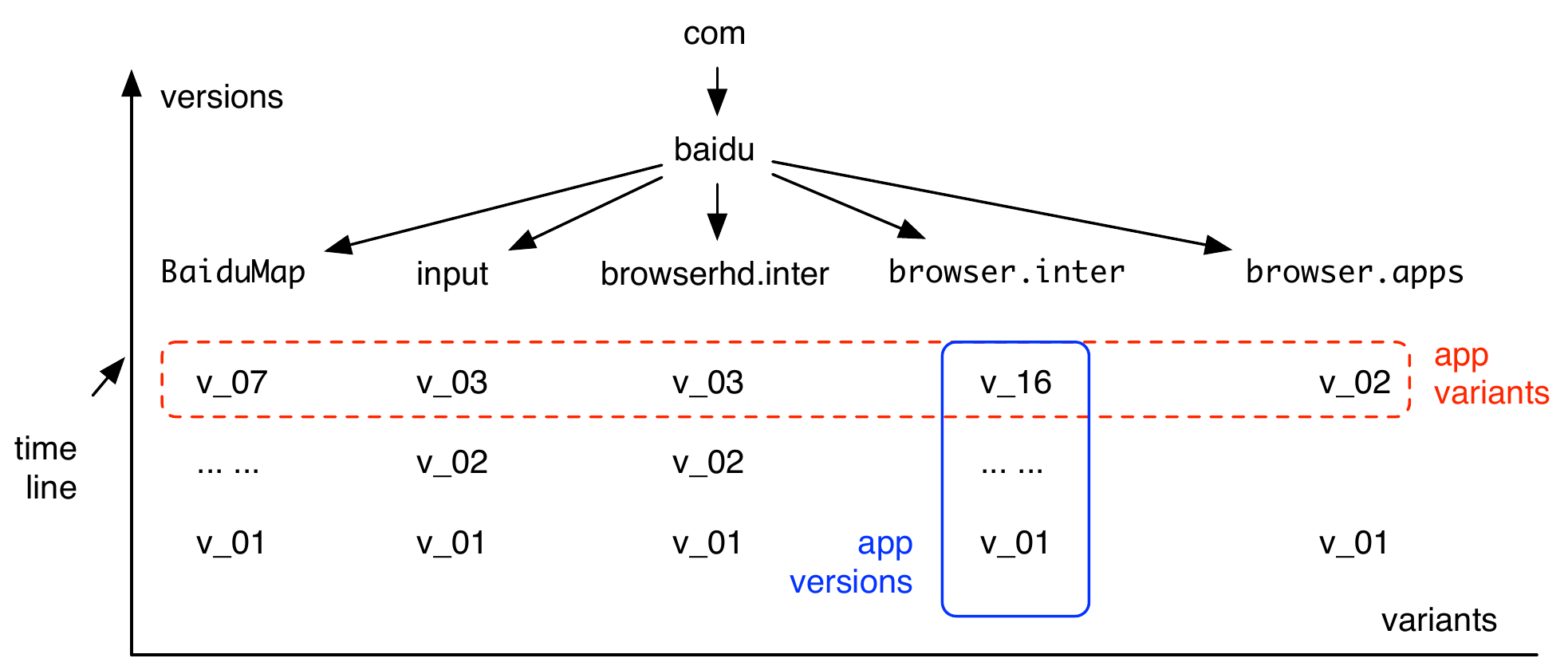}
    \caption{A Simplified Example Showing how App Variants and Versions are Selected.}
    \label{fig:variants}
\end{figure}

Fig.~\ref{fig:variants} presents a tree model to select app variants from a set of to-be mined Android apps, where each non-leaf node is represented by a package segment (e.g., Baidu) while leaf node is represented by the remaining package segments (e.g., BaiduMap).
A branch from the root node (i.e., com) to a leaf node (e.g., BaiduMap) represents a unique package name (i.e., the Baidu Map).
Because each Android app can have different versions, e.g., each update will result in a version, each leaf node has further been affiliated with a list of meta-data of app versions.
The time line of the vertical axis shows that the affiliated list is ranked through times and the apps it contains are actually different versions of the app indicated by the leaf node. 
Given a time point, the tree model also gives a way to identify family variants. 
For example, as shown in the red dashed rectangle, given the latest time point, we are able to collect a set of variants for company com.baidu.

In this research direction, based on a clustering-based approach, we have collected in total 75,963 families of apps: The median number of variants in a family is three and 760 collected families have over 100 variants.
Through a preliminary study on the collected app families, we have identified several reuse cases adopted by Android app variants, including library reuse, automated app generation, content-driven variants and device-driven variants~\cite{li2016mining}.
With an advanced study, we believe that more reuse cases, including semantic reuses, can be identified on top of our collect app family variants. 

\textbf{Pairwise Similarity Analysis.}
We present a research framework called SimiDroid~\cite{li2017simidroid} that supports multi-level pairwise similarity comparison of Android apps, aiming at supporting the understanding of similarities or changes among app versions and among repackaged apps.
SimiDroid is designed as a plugin-based framework that has already integrated various comparison methods such as code-based or resource-based comparisons.
In addition to detect similar Android apps, we also perform a number of case studies on AndroZoo apps to demonstrate the suitability of SimiDroid in providing explanation hints for different usage scenarios.
For example, we have leveraged SimiDroid to check and validate the hypothesis of multi-generation repackaging, where an original app identified in a repackaging pair is actually a repackaged app from a prior repackaging generation~\cite{li2017the}.

\begin{figure}[!h]
    \centering
    \includegraphics[width=0.8\linewidth]{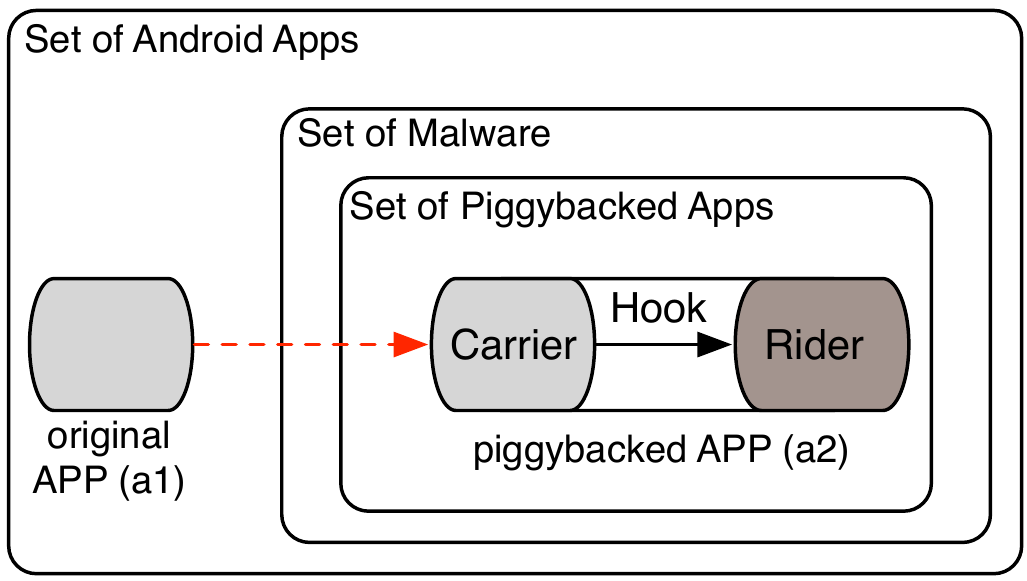}
    \caption{Piggybacking Terminology.}
    \label{fig:piggybacking}
\end{figure}

\textbf{Piggybacking Behaviour Understanding.}
Fig.~\ref{fig:piggybacking} presents some basic terms related to Android app piggybacking.
Basically, the working process of piggybacking is like this: 
Given an original Android app (referred to as carrier), attackers first unpack it and then modify its code by injecting some additional code (referred to as rider), and finally re-pack it back to a new app version (referred to as piggybacked app).
The injected code will be triggered thanks to the so-called hooks, which connect the execution of carrier code to rider code.

Despite many research works have been conducted in the literature to detect piggybacked apps, the literature lacks a comprehensive study on the behaviour of piggybacking.
To this end, we construct a benchmark set of piggybacked app pairs (through pairwise similarity analysis) and investigate the characteristics of malicious piggybacked apps in comparison with their original counterparts.
Thanks to these comparisons, we have eventually observed many interesting findings on the piggybacking process~\cite{li2017understanding,li2017understanding2}.
The findings are summarised as follows:

\begin{enumerate}
\item The realisation of malicious behaviour is often accompanied by manipulation of app resource files.
\item Piggybacking changes app behaviour mostly by tampering with carrier code.
\item Piggybacked apps are potentially built in batches.
\item Piggybacking often asks for new permissions to allow the realisation of malicious behaviour.
\item Piggybacking may recurrently request some specific permissions that are less requested by non-piggybacked apps.
\item Piggybacking is probably largely automated.
\item Piggybacking may overly request permissions that have been already declared by their original apps.
\item Piggybacking may introduce new user interfaces, implement new receivers and services, but will not add new database structures.
\item Piggybacking often consists in injecting a component that offers the same capabilities (i.e., Action, Category, etc.) as an existing component in the original app.
\item Piggybacking may change the launcher component so as to trigger the execution of rider code.
\item Piggybacking is often characterised by a naming mismatch between existing and newly injected components.
\item Piggybacking generally connects the malicious payloads to the benign carrier code via a single method call, making it possible to automatically locate grafted malicious payloads from piggybacked malicious apps.
\item Piggybacking hooks are generally placed within library code rather than in core app code.
\item Piggybacking often reuses the to-be injected malicious payloads.
\item Piggybacking adds code which performs sensitive actions, often without referring to device users.
\item Piggybacking operations are distributed over well-known malicious behaviour types.
\item Piggybacking increasingly hides malicious actions via uses of reflection and dynamic class loading.
\item Piggybacking complicates app's overall call graph, while rider code can even largely exceed in size the carrier code.
\item Piggybacking is seldom conducted by authors of benign apps.
\item Piggybacking code brings more execution paths where sensitive data can be leaked.
\end{enumerate}

\subsection{Malware Analysis}

\textbf{Potential Component Leaks.}
Potential Component Leaks are such attacks that leak sensitive information through known ICC vulnerabilities such as Activity Hijacking Attack and Broadcast Injection Attack~\cite{octeau2013effective}.
In this work, we have defined two types potential component leaks: 
1) potential passive component leak, which leaks everything a component receives from other components; and 2) potential active component leak, which attempts to send sensitive information to other components.
In order to detect the aforementioned attacks, we present a prototype tool named PCLeaks, which based on ICC vulnerabilities to perform data-flow analysis on Android apps to pinpoint potential component leaks that could potentially be exploited by other components (or apps)~\cite{li2014automatically}.

\textbf{ML-based Malware Detection.}
As a follow-up work of PCLeaks, we find that potential component leaks are common in Android apps and that malicious apps have manipulated significantly more potential component leaks than benign apps.
This evidence makes potential component leaks perfect candidates of features for machine learning (ML) based malware detection.
Towards verifying this hypothesis, we take potential component leaks as features to train several classification models (with different settings such as different benign/malware ration, different ML algorithm, etc.) and perform 10-fold cross-validation to justify the ability to identify malicious Android apps.
Our experimental validations show high performance for identifying malware, demonstrating that potential component leaks are useful for discriminating malicious from benign apps~\cite{li2015potential}.

\textbf{Topic-Specific Data-Flow Analysis.}
State-of-the-art work has shown that both app descriptions~\cite{gorla2014checking} and sensitive data-flows~\cite{avdiienko2015mining} in standalone are capable of discriminating malicious from benign apps.
In this work, we take both app descriptions (i.e., indicative of app topics) and sensitive data-flows (i.e., functionality implemented) into consideration for discriminating malware from benign apps.
At the beginning, we leverage adaptive LDA with GA, an advanced topic model, to cluster apps different categories based on their descriptions.
Then, we use information gain ratio of sensitive data-flows to build so-called ``topic-specific data-flow signatures''.
Finally, we leverage those signatures to characterise malicious Android apps.
Our experiments on 3,691 benign and 1,612 malicious apps demonstrate that topic-specific data-flow signatures are useful and effective in highlighting malicious app behaviour~\cite{yang2017characterizing}.

\textbf{Malicious Payload Identification.}
As shown in Fig.~\ref{fig:piggybacking}, the malicious payloads of piggybacked apps are usually triggered by a small piece of code called hooks.
If we are able to locate such hooks, we can accurately locate the malicious payloads, and thus reducing the examination space for security analysts to understand the malicious behaviour.
To this end, we propose in this work a tool-based approach called HookRanker~\cite{li2017automatically}, which provides rank lists of potential malicious packages based on the way malware behaviour are revealed.
With experiments on a ground truth of piggybacked Android apps, we demonstrate that HookRanker is helpful for locating malicious packages of piggybacked Android malware.

\section{Leveraging AndroZoo}
\label{sec:leveraging}
Our AndroZoo dataset, as shown in the previous section, has already been used by us to conduct various Android-based research in the field of Security and Software Engineering~\cite{li2017mining}.
It has also been frequently leveraged by our worldwide fellow researchers to conduct world-class exploration~\cite{hecht2015tracking,calciati2017apps,li2017android}.
In general, we believe our dataset could be leveraged for investigating in the fields of Code Recommendation, large scale studies on API usage and adoption, coding patterns, repackaging detection, Library detection and popularity analysis, Obfuscation techniques, Malware analysis, application similarity, etc.
Our dataset, thanks to its long-term coverage, is particularly well suited to enable evolution studies. Indeed, our dataset, with the latest snapshot, contains more than 450 thousand apps with over 28 app lineages for which each of them has 10 or more versions.

\textbf{Access Conditions.}
We make our dataset available to the research community. Given the lack of a clear, universal copyright exemption for Research, we request that researchers willing to access this dataset: 
\begin{itemize}
\item Evaluate the legal situation of downloading and working on copyrighted applications with regards to their situation (local laws, host institution policy, etc);
\item Do not, in general, redistribute the data;
\item Do not, in particular, make a commercial usage of this data; 
\item Act responsibly with this data, notably with regards to the maliciousness of many apps;
\item Get a faculty, or someone in a permanent position, to agree and commit to those conditions.
\end{itemize}

We politely ask that the origin of the dataset be acknowledged, and we hope that researchers will make available the lists of apps used in their publications to ensure their experiments are reproducible.

\section{Conclusion}
\label{sec:conclusion}
We have presented the AndroZoo dataset of millions of Android apps collected from various data sources and their metadata collected via various means.
So far, we collected over five million apps and over 30 artefacts of metadata associated with them.
We make these datasets readily available to the community to contribute to more reliable and reproducible studies based on a large-scale, representative, and up-to-date samples.

\balance
\bibliographystyle{unsrt}
\bibliography{androzoo}

\begin{thebibliography}{10}

\bibitem{li2017understanding}
Li~Li, Daoyuan Li, Tegawend{\'e}~F Bissyand{\'e}, Jacques Klein, Yves Le~Traon,
  David Lo, and Lorenzo Cavallaro.
\newblock Understanding android app piggybacking: A systematic study of
  malicious code grafting.
\newblock {\em IEEE Transactions on Information Forensics \& Security (TIFS)},
  2017.

\bibitem{li2016accessing}
Li~Li, Tegawend{\'e}~F Bissyand{\'e}, Yves Le~Traon, and Jacques Klein.
\newblock Accessing inaccessible android apis: An empirical study.
\newblock In {\em The 32nd International Conference on Software Maintenance and
  Evolution (ICSME 2016)}, 2016.

\bibitem{felt2011android}
Adrienne~Porter Felt, Erika Chin, Steve Hanna, Dawn Song, and David Wagner.
\newblock Android permissions demystified.
\newblock In {\em Proceedings of the 18th ACM conference on Computer and
  communications security}, pages 627--638. ACM, 2011.

\bibitem{backes2016demystifying}
Michael Backes, Sven Bugiel, Erik Derr, Patrick~D McDaniel, Damien Octeau, and
  Sebastian Weisgerber.
\newblock On demystifying the android application framework: Re-visiting
  android permission specification analysis.
\newblock In {\em USENIX Security Symposium}, pages 1101--1118, 2016.

\bibitem{egele2013empirical}
Manuel Egele, David Brumley, Yanick Fratantonio, and Christopher Kruegel.
\newblock An empirical study of cryptographic misuse in android applications.
\newblock In {\em Proceedings of the 2013 ACM SIGSAC conference on Computer \&
  communications security}, pages 73--84. ACM, 2013.

\bibitem{gorla2014checking}
Alessandra Gorla, Ilaria Tavecchia, Florian Gross, and Andreas Zeller.
\newblock Checking app behavior against app descriptions.
\newblock In {\em Proceedings of the 36th International Conference on Software
  Engineering}, pages 1025--1035. ACM, 2014.

\bibitem{li2017static}
Li~Li, Tegawend{\'e}~F Bissyand{\'e}, Mike Papadakis, Siegfried Rasthofer,
  Alexandre Bartel, Damien Octeau, Jacques Klein, and Yves Le~Traon.
\newblock Static analysis of android apps: A systematic literature review.
\newblock {\em Information and Software Technology}, 2017.

\bibitem{li2015apkcombiner}
Li~Li, Alexandre Bartel, Tegawend{\'e}~F Bissyand{\'e}, Jacques Klein, and Yves
  Le~Traon.
\newblock {ApkCombiner: Combining Multiple Android Apps to Support Inter-App
  Analysis}.
\newblock In {\em Proceedings of the 30th IFIP International Conference on ICT
  Systems Security and Privacy Protection (SEC 2015)}, 2015.

\bibitem{li2015iccta}
Li~Li, Alexandre Bartel, Tegawend{\'e}~F Bissyand{\'e}, Jacques Klein, Yves
  Le~Traon, Steven Arzt, Siegfried Rasthofer, Eric Bodden, Damien Octeau, and
  Patrick Mcdaniel.
\newblock {IccTA: Detecting Inter-Component Privacy Leaks in Android Apps}.
\newblock In {\em Proceedings of the 37th International Conference on Software
  Engineering (ICSE 2015)}, 2015.

\bibitem{li2014know}
Li~Li, Alexandre Bartel, Jacques Klein, Yves~Le Traon, Steven Arzt, Siegfried
  Rasthofer, Eric Bodden, Damien Octeau, and Patrick Mcdaniel.
\newblock I know what leaked in your pocket: uncovering privacy leaks on
  android apps with static taint analysis.
\newblock {\em arXiv preprint arXiv:1404.7431}, 2014.

\bibitem{arzt2014flowdroid}
Steven Arzt, Siegfried Rasthofer, Christian Fritz, Eric Bodden, Alexandre
  Bartel, Jacques Klein, Yves Le~Traon, Damien Octeau, and Patrick McDaniel.
\newblock Flowdroid: Precise context, flow, field, object-sensitive and
  lifecycle-aware taint analysis for android apps.
\newblock {\em Acm Sigplan Notices}, 49(6):259--269, 2014.

\bibitem{octeau2015composite}
Damien Octeau, Daniel Luchaup, Matthew Dering, Somesh Jha, and Patrick
  McDaniel.
\newblock Composite constant propagation: Application to android
  inter-component communication analysis.
\newblock In {\em Proceedings of the 37th International Conference on Software
  Engineering-Volume 1}, pages 77--88. IEEE Press, 2015.

\bibitem{octeau2016combining}
Damien Octeau, Somesh Jha, Matthew Dering, Patrick Mcdaniel, Alexandre Bartel,
  Li~Li, Jacques Klein, and Yves Le~Traon.
\newblock Combining static analysis with probabilistic models to enable
  market-scale android inter-component analysis.
\newblock In {\em Proceedings of the 43th Symposium on Principles of
  Programming Languages (POPL 2016)}, 2016.

\bibitem{li2016droidra}
Li~Li, Tegawend{\'e}~F Bissyand{\'e}, Damien Octeau, and Jacques Klein.
\newblock Droidra: Taming reflection to support whole-program analysis of
  android apps.
\newblock In {\em The 2016 International Symposium on Software Testing and
  Analysis (ISSTA 2016)}, 2016.

\bibitem{li2016reflection}
Li~Li, Tegawend{\'e}~F Bissyand{\'e}, Damien Octeau, and Jacques Klein.
\newblock Reflection-aware static analysis of android apps.
\newblock In {\em The 31st IEEE/ACM International Conference on Automated
  Software Engineering, Demo Track (ASE 2016)}, 2016.

\bibitem{li2016parameter}
Li~Li, Tegawend{\'e}~F Bissyand{\'e}, Jacques Klein, and Yves Le~Traon.
\newblock {Parameter Values of Android APIs: A Preliminary Study on 100,000
  Apps}.
\newblock In {\em Proceedings of the 23rd IEEE International Conference on
  Software Analysis, Evolution, and Reengineering (SANER 2016)}, 2016.

\bibitem{wang2015wukong}
Haoyu Wang, Yao Guo, Ziang Ma, and Xiangqun Chen.
\newblock Wukong: a scalable and accurate two-phase approach to android app
  clone detection.
\newblock In {\em Proceedings of the 2015 International Symposium on Software
  Testing and Analysis}, pages 71--82. ACM, 2015.

\bibitem{avdiienko2015mining}
Vitalii Avdiienko, Konstantin Kuznetsov, Alessandra Gorla, Andreas Zeller,
  Steven Arzt, Siegfried Rasthofer, and Eric Bodden.
\newblock Mining apps for abnormal usage of sensitive data.
\newblock In {\em Proceedings of the 37th International Conference on Software
  Engineering-Volume 1}, pages 426--436. IEEE Press, 2015.

\bibitem{gibler2013adrob}
Clint Gibler, Ryan Stevens, Jonathan Crussell, Hao Chen, Hui Zang, and Heesook
  Choi.
\newblock Adrob: Examining the landscape and impact of android application
  plagiarism.
\newblock In {\em Proceeding of the 11th annual international conference on
  Mobile systems, applications, and services}, pages 431--444. ACM, 2013.

\bibitem{grace2012unsafe}
Michael~C Grace, Wu~Zhou, Xuxian Jiang, and Ahmad-Reza Sadeghi.
\newblock Unsafe exposure analysis of mobile in-app advertisements.
\newblock In {\em Proceedings of the fifth ACM conference on Security and
  Privacy in Wireless and Mobile Networks}, pages 101--112. ACM, 2012.

\bibitem{martinez2017bottom}
Jabier Martinez, Tewfik Ziadi, Tegawend{\'e}~F Bissyand{\'e}, Jacques Klein,
  and Yves~Le Traon.
\newblock Bottom-up technologies for reuse: automated extractive adoption of
  software product lines.
\newblock In {\em Proceedings of the 39th International Conference on Software
  Engineering Companion}, pages 67--70. IEEE Press, 2017.

\bibitem{li2016mining}
Li~Li, Jabier Martinez, Tewfik Ziadi, Tegawend{\'e}~F Bissyand{\'e}, Jacques
  Klein, and Yves Le~Traon.
\newblock Mining families of android applications for extractive spl adoption.
\newblock In {\em The 20th International Systems and Software Product Line
  Conference (SPLC 2016)}, 2016.

\bibitem{li2017simidroid}
Li~Li, Tegawend{\'e}~F Bissyand{\'e}, and Jacques Klein.
\newblock Simidroid: Identifying and explaining similarities in android apps.
\newblock In {\em The 16th IEEE International Conference On Trust, Security And
  Privacy In Computing And Communications (TrustCom)}, 2017.

\bibitem{li2017the}
Li~Li, Tegawend{\'e}~F Bissyand{\'e}, Alexandre Bartel, Jacques Klein, and Yves
  Le~Traon.
\newblock The multi-generation repackaging hypothesis.
\newblock In {\em The 39th International Conference on Software Engineering,
  Poster Track (ICSE 2017)}, 2017.

\bibitem{li2017understanding2}
Li~Li, Daoyuan Li, Tegawend{\'e}~F Bissyand{\'e}, Jacques Klein, Yves Le~Traon,
  David Lo, and Lorenzo Cavallaro.
\newblock Understanding android app piggybacking.
\newblock In {\em The 39th International Conference on Software Engineering,
  Poster Track (ICSE 2017)}, 2017.

\bibitem{octeau2013effective}
Damien Octeau, Patrick McDaniel, Somesh Jha, Alexandre Bartel, Eric Bodden,
  Jacques Klein, and Yves Le~Traon.
\newblock Effective inter-component communication mapping in android with
  epicc: An essential step towards holistic security analysis.
\newblock In {\em USENIX Security}, 2013.

\bibitem{li2014automatically}
Li~Li, Alexandre Bartel, Jacques Klein, and Yves Le~Traon.
\newblock Automatically exploiting potential component leaks in android
  applications.
\newblock In {\em Proceedings of the 13th International Conference on Trust,
  Security and Privacy in Computing and Communications (TrustCom 2014)}, 2014.

\bibitem{li2015potential}
Li~Li, Kevin Allix, Daoyuan Li, Alexandre Bartel, Tegawend{\'e}~F
  Bissyand{\'e}, and Jacques Klein.
\newblock {Potential Component Leaks in Android Apps: An Investigation into a
  new Feature Set for Malware Detection}.
\newblock In {\em The 2015 IEEE International Conference on Software Quality,
  Reliability \& Security (QRS)}, 2015.

\bibitem{yang2017characterizing}
Xinli Yang, David Lo, Li~Li, Xin Xia, Tegawend{\'e}~F Bissyand{\'e}, and
  Jacques Klein.
\newblock Characterizing malicious android apps by mining topic-specific data
  flow signatures.
\newblock {\em Information and Software Technology}, 2017.

\bibitem{li2017automatically}
Li~Li, Daoyuan Li, Tegawend{\'e}~F Bissyand{\'e}, Jacques Klein, Haipeng Cai,
  David Lo, and Yves Le~Traon.
\newblock Automatically locating malicious packages in piggybacked android
  apps.
\newblock In {\em The 4th IEEE/ACM International Conference on Mobile Software
  Engineering and Systems (MobileSoft 2017)}, 2017.

\bibitem{li2017mining}
Li~Li.
\newblock Mining androzoo: A retrospect.
\newblock In {\em The International Conference on Software Maintenance and
  Evolution}, 2017.

\bibitem{hecht2015tracking}
Geoffrey Hecht, Omar Benomar, Romain Rouvoy, Naouel Moha, and Laurence Duchien.
\newblock Tracking the software quality of android applications along their
  evolution (t).
\newblock In {\em Automated Software Engineering (ASE), 2015 30th IEEE/ACM
  International Conference on}, pages 236--247. IEEE, 2015.

\bibitem{calciati2017apps}
Paolo Calciati and Alessandra Gorla.
\newblock How do apps evolve in their permission requests?: a preliminary
  study.
\newblock In {\em Proceedings of the 14th International Conference on Mining
  Software Repositories}, pages 37--41. IEEE Press, 2017.

\bibitem{li2017android}
Yuping Li, Jiyong Jang, Xin Hu, and Xinming Ou.
\newblock Android malware clustering through malicious payload mining.
\newblock {\em arXiv preprint arXiv:1707.04795}, 2017.

\end{thebibliography}

\end{document}